\documentclass[twocolumn,superscriptaddress,prl]{revtex4}
\usepackage{amsmath}
\usepackage{graphicx}
\usepackage{xcolor} 
\usepackage[normalem]{ulem} 

\newcommand{\SB}{\color{black}}
\newcommand{\stkout}[1]{\ifmmode\text{\sout{\ensuremath{#1}}}\else\sout{#1}\fi}
\newcommand{\md}{\mathrm d}

\newcommand{\Ntot}{N_{\rm tot}}
\newcommand{\Nub}{N_{\rm UB}}
\newcommand{\Nb}{N_{\rm B}}
\newcommand{\muI}{\mu_{\rm i}}
\newcommand{\muF}{\mu_{\rm f}}

\newcommand{\kT}{k_{\rm B} T}
\DeclareMathOperator{\sech}{sech}

\begin{document}

\title{Steps minimize dissipation in rapidly driven stochastic systems}

\author{Steven Blaber}
\email{sblaber@sfu.ca}
\affiliation{Dept.~of Physics, Simon Fraser University, Burnaby, British Columbia V5A 1S6, Canada}
\author{Miranda D.\ Louwerse}
\affiliation{Dept.~of Chemistry, Simon Fraser University, Burnaby, British Columbia V5A 1S6, Canada}
\author{David A.\ Sivak}
\email{dsivak@sfu.ca}
\affiliation{Dept.~of Physics, Simon Fraser University, Burnaby, British Columbia V5A 1S6, Canada}

\begin{abstract}
Micro- and nano-scale systems driven by rapid changes in control parameters (control protocols) dissipate significant energy. In the fast-protocol limit, we find that protocols that minimize dissipation at fixed duration are universally given by a two-step process, jumping to and from a point that balances jump size with fast relaxation. Jump protocols could be exploited by molecular machines or thermodynamic computing to improve energetic efficiency, and implemented in nonequilibrium free-energy estimation to improve accuracy.
\end{abstract}

\maketitle

The birth of thermodynamics as a modern science can be traced to Sadi Carnot's study of the design principles for energetically efficient heat engines described in \emph{Reflections on the Motive Power of Fire}~\cite{Carnot1960}. In classical thermodynamics, minimum-dissipation protocols are important in the design of macroscopic heat engines describing, for example, adiabatic (no heat loss) and quasistatic (infinitely slow) compression of gas by a piston. Nearly 200 years later, the field of stochastic thermodynamics~\cite{Jarzynski2011,Seifert2012} similarly studies the design principles governing the ability to dynamically vary control parameters and perform work at minimum energetic cost (minimum dissipation), but now in micro- and nano-scale fluctuating systems. Minimum-dissipation protocols inform our understanding of the design principles of biological molecular machines~\cite{Brown2017,Brown2019} and are of practical use to single-molecule experiments~\cite{Tafoya2019}, free-energy estimation~\cite{Shenfeld2009,Geiger2010,Blaber2020,Chodera2011,Minh2009optimal}, and thermodynamic computing~\cite{Conte2019,Proesmans2020}.

In contrast to macroscopic systems that are well described by averages of thermodynamic quantities, the stochastic fluctuations in microscopic systems are large relative to the mean and cannot be ignored. The work done on a stochastic system by a control protocol is a fluctuating quantity that depends on the entire protocol history, making it particularly difficult to optimize. General optimization requires minimizing over all possible paths through control-parameter space, which cannot be solved in general~\cite{Schmiedl2007}. Despite advances relating optimal-control to optimal-transport theory, even numerical optimizations are still limited to relatively simple systems~\cite{Aurell2011}. 

Although general solutions are intractable, we have gained considerable insight into minimum-dissipation protocols by considering simple systems. For example, Schmiedl and Seifert~\cite{Schmiedl2007} showed that for a Brownian particle diffusing in a harmonic potential with time-dependent minimum or stiffness, minimum-dissipation protocols exhibit jump discontinuities. It was posited that jumps in control parameters are a general feature of minimum-dissipation protocols, and they have since been observed in a number of different systems~\cite{Gomez2008,Then2008,Esposito2010}.

More general insight can be gained by approximating the mean work in relevant limits. For slow protocols, linear-response theory yields an approximation for the mean work, from which the approximate minimum-dissipation protocol can be calculated~\cite{OptimalPaths}. Despite its success, the linear-response formalism relies on near-equilibrium approximations that break down for fast protocols, miss key features of the exact minimum-dissipation protocol (e.g., jumps in control parameters), and for short duration can perform worse than naive (constant-velocity) protocols~\cite{Blaber2020Skewed}.

\begin{figure}
    \includegraphics[width=\linewidth]{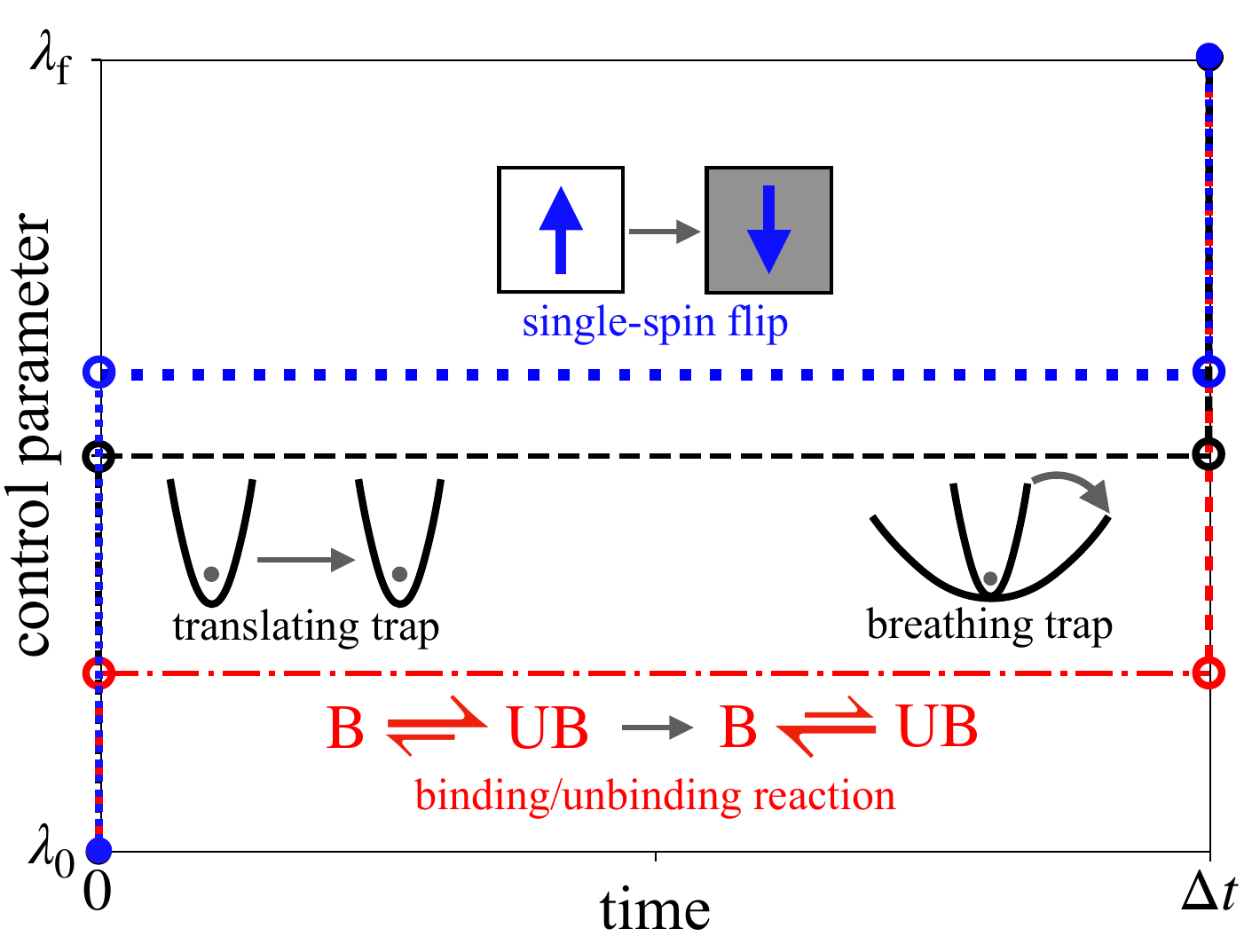}
	\caption{Short-time efficient protocols (STEPs) for a single spin in a time-dependent magnetic field (blue dots), Brownian particle in a translating harmonic potential (black dashed), Brownian particle in a harmonic potential with time-dependent stiffness (black dashed), and a two-state binding/unbinding reaction system with variable binding and unbinding rates controlled by the chemical-potential difference (red dash-dotted).}
	\label{STEP protocol}
\end{figure}

\begin{figure}
	\includegraphics[width=\linewidth]{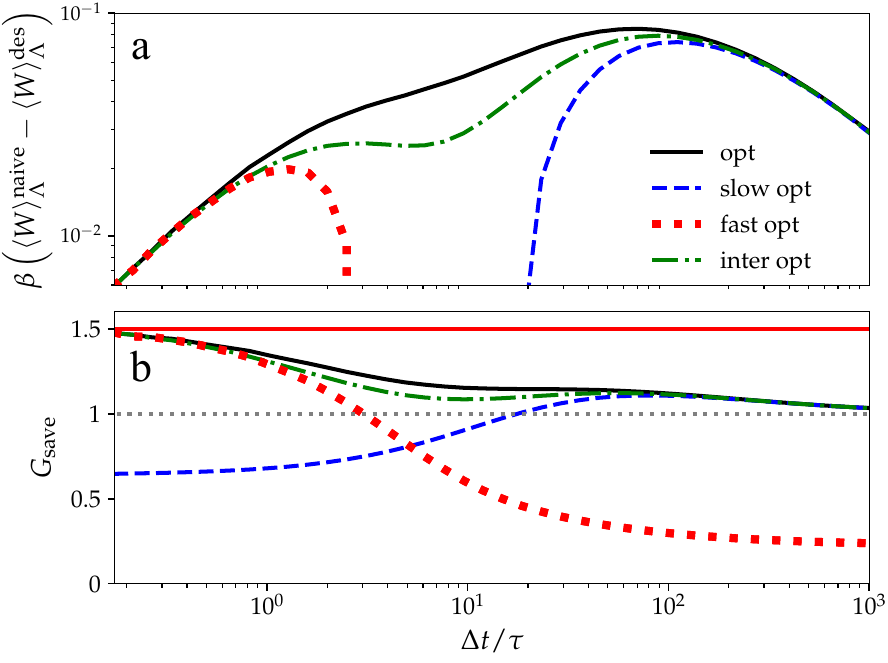}
	\caption{Benefit in the breathing harmonic trap from designed protocols relative to the naive (constant-velocity) protocol, as a function of the duration $\Delta t$ scaled by the fastest integral relaxation time $\tau$. The different designed (``des'') protocols include the exact optimal~\cite{Schmiedl2007} (``opt'', solid black), linear-response optimized (``slow opt'', dashed blue), STEP (``fast opt'', red dots), and interpolated optimal protocol (``inter opt'', dash-dotted green). (a) Difference between the work done by the naive (constant-velocity) and designed protocols. (b) Gain $G_{\rm save}\equiv {\langle W_{\rm save} \rangle_{\Lambda}^{\rm des}} /{\langle W_{\rm save} \rangle_{\Lambda}^{\rm naive}}$ in saved work. Solid red line in (b) denotes the short-duration limit~\eqref{Fast_Ratio_Appx}. Control-parameter endpoints satisfy $\lambda_{\rm i}/\lambda_{\rm f} = 16$, and the interpolated protocol uses $\alpha = 1$ and fastest integral relaxation time $\tau = 1/(2\lambda_{\rm i})$~\cite{Blaber2020Skewed}.}
	\label{BHT_Work_Diff}
\end{figure}

\begin{figure}
    \centering
	\includegraphics[width=\linewidth]{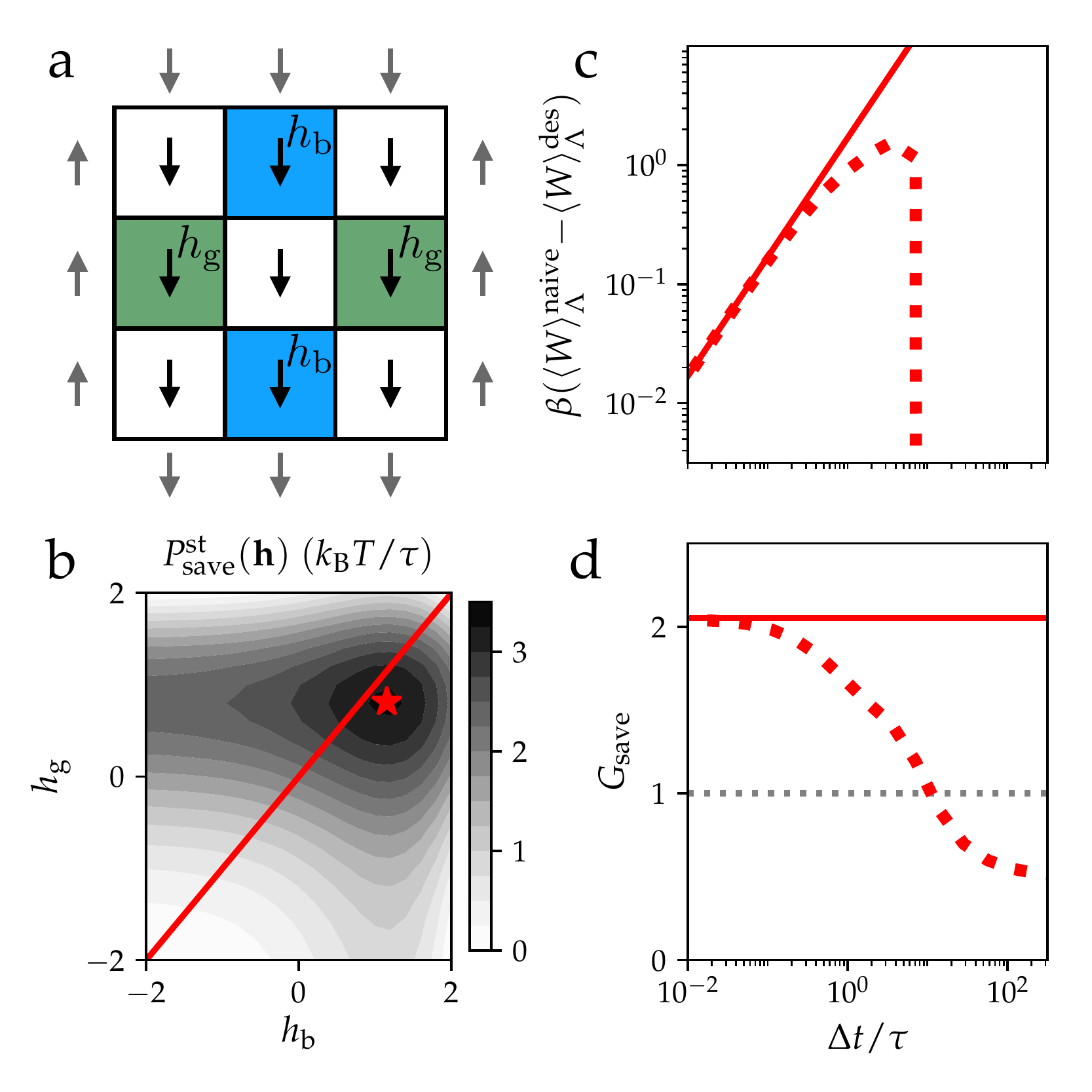}
	\caption{a) Nine-spin ferromagnetic Ising model (internal black spins) with fixed boundary conditions (external gray spins). The multi-dimensional control parameter is two external magnetic fields, $h_{\rm b}$ (blue) applied to horizontal-edge spins and $h_{\rm g}$ (green) applied to vertical-edge spins. b) Short-time power savings~\eqref{eq:power_savings} as function of control parameters ($h_{\rm b}$,$h_{\rm g}$). Red line: naive protocol; red star: $\boldsymbol{h}^{\rm{STEP}}$~\eqref{Euler Lagrange 1}. c) Work difference between designed and naive protocols (dotted red), and its short-duration approximation~\eqref{Excess work approx 2} (solid red). d) Gain $G_{\rm save}\equiv {\langle W_{\rm save} \rangle_{\Lambda}^{\rm des}} /{\langle W_{\rm save} \rangle_{\Lambda}^{\rm naive}}$ in saved work for multi-dimensional STEP relative to naive (dotted red), and its short-duration limit~\eqref{Fast_Ratio_Appx} (solid red). Control-parameter endpoints are $\boldsymbol{h}_{\rm i} = (-2,-2)$ and $\boldsymbol{h}_{\rm f} = (2,2)$, with duration $\Delta t$ and fastest relaxation time $\tau=N/k_0$, for $N=9$ spins and single-spin flip attempt rate $k_0$.}
	\label{MultiD_Ising}
\end{figure}

While minimum-dissipation protocols for slowly driven systems are relatively well understood, comparatively little is known about rapidly driven systems. In this work we focus on fast protocols and find a universal design principle: the minimum-dissipation protocol consists of jumps at the beginning and end of the protocol, spending the entire duration at the control-parameter value that optimally balances the \emph{initial force-relaxation rate} (IFRR)~\eqref{Rate Function} with the jump size~\eqref{Euler Lagrange 1}. Our results are physically intuitive, apply to a wide range of {\SB stochastic} systems, and generalize easily to multidimensional control. To illustrate this, we calculate the minimum-dissipation protocols in a diverse set of systems described by Fokker-Planck or master-equation dynamics with single- (Fig.~\ref{STEP protocol}) or multi-dimensional control (Fig.~\ref{MultiD_Ising}). Combining our results with known minimum-dissipation protocols in the slow limit~\cite{OptimalPaths}, we demonstrate that a simple interpolation scheme produces protocols that reduce dissipation at all speeds (Fig.~\ref{BHT_Work_Diff}).

\emph{Derivation}.---Consider a {\SB stochastic} thermodynamic system described by dynamics of the form
\begin{align}
\frac{\partial p_{\Lambda}(x,t)}{\partial t}  = L[x,\boldsymbol{\lambda}(t)]\, p_{\Lambda}(x,t) \ ,
\label{Exact probability evolution}
\end{align}
where $p_{\Lambda}(x,t)$ is the probability distribution over microstates $x$ at time $t$ given the control protocol $\Lambda$, and $L[x,\boldsymbol{\lambda}(t)]$ is the operator describing the system's time evolution. $L$ is the drift/diffusion operator for Fokker-Planck and the transition-rate matrix for master-equation dynamics. The system is in contact with a heat bath at temperature $T$ such that the equilibrium probability distribution over microstates $x$ at fixed control parameters $\boldsymbol{\lambda}$  is $\pi(x|\boldsymbol{\lambda}) \equiv \exp\{\beta[F(\boldsymbol{\lambda})-U(x,\boldsymbol{\lambda})]\}$, for internal energy $U(x,\boldsymbol{\lambda})$ and free energy $F(\boldsymbol{\lambda}) \equiv -k_{\rm B}T\, \ln\sum_{x}\exp[-\beta U(x,\boldsymbol{\lambda})] $, where $\beta \equiv (k_{\rm B}T)$ for Boltzmann's constant $k_{\rm B}$. The average excess work $W_{\rm ex} \equiv W-\Delta F$ by an external agent changing control parameters $\boldsymbol{\lambda}$ according to protocol $\Lambda$ is
\begin{align}
\langle W_{\rm ex} \rangle_{\Lambda} = -\int_{0}^{\Delta t }\md t \,\frac{\md \boldsymbol{\lambda}^{\rm T}}{\md t} \langle \delta {\bf f}(t) \rangle_{\Lambda} \ ,
\label{work def}
\end{align}
where a bold symbol denotes a column vector and superscript ${\rm T}$ the transpose.  ${\bf f} \equiv -\partial U/\partial\boldsymbol{\lambda}$ are the forces conjugate to the control parameters, and $\delta {\bf f} \equiv {\bf f} - \langle {\bf f} \rangle_{\rm eq}$ the deviations from the equilibrium averages. Angle brackets $\langle \cdots\rangle_{\Lambda}$ denote a nonequilibrium ensemble average over the control-parameter protocol $\Lambda$. Here we hold fixed the initial ($\boldsymbol{\lambda}_{\rm i}$) and final ($\boldsymbol{\lambda}_{\rm f}$) control parameters, consistent with nonequilibrium free-energy estimation~\cite{Ritort2002,Gore2003,Shenfeld2009,Geiger2010,Kim2012,Blaber2020Skewed,Schindler2020,Aldeghi2018,Kuhn2017,Ciordia2016,Wang2015,Gapsys2015,Chodera2011,Minh2009optimal} but distinct from optimizations that constrain the initial and final probability distributions~\cite{Proesmans2020,Zhang2020}.

If the total duration $\Delta t$ is short compared to the system's natural relaxation time $\tau$ (a fast protocol), expanding the probability distribution in $\Delta t/\tau$ around an initial equilibrium distribution gives
\begin{align}
p_{\Lambda}(x,s)= \pi(x|\boldsymbol{\lambda}_{\rm i})+ p^{1}_{\Lambda}(x,s) \frac{\Delta t}{\tau} + \mathcal{O}\left[ \left(\frac{\Delta t}{\tau}\right)^2\right] \ ,
\label{Probability expansion}
\end{align}
for $s \equiv t/\Delta t$ and first-order coefficient $p^{1}_{\Lambda}(x,s)$. Plugging \eqref{Probability expansion} up to $\mathcal{O}(\Delta t/\tau)$ into \eqref{Exact probability evolution} gives
\begin{align}
\frac{\partial p^{1}_{\Lambda}(x,s)}{\partial s} \approx \mathcal{L}[x,\boldsymbol{\lambda}(s)]\, \pi(x|\boldsymbol{\lambda}_{\rm i}) \ ,
\end{align}
with $\mathcal{L} \equiv \tau L$ the dimensionless time-evolution operator. Solving for $p^{1}_{\Lambda}(x,s)$ and substituting into \eqref{Probability expansion} yields
\begin{align}
p_{\Lambda}(x,s)\approx \pi(x|\boldsymbol{\lambda}_{\rm i})+\frac{\Delta t}{\tau}\int_{0}^{s}\md s'\mathcal{L}[x,\boldsymbol{\lambda}(s')]\, \pi(x|\boldsymbol{\lambda}_{\rm i}) \ .
\label{Short time prop}
\end{align}
Multiplying by conjugate forces ${\bf f}$, integrating over microstates $x$, and changing the time variable back to $t$ gives
\begin{align}
\langle {\bf f}(t) \rangle_{\Lambda} \approx \langle {\bf f} \rangle_{\boldsymbol{\lambda}_{\rm i}}+\int_{0}^{t}\md t' \, {\bf R}_{\boldsymbol{\lambda}_{\rm i}}[\boldsymbol{\lambda}(t')] \ ,
\label{Short time force}
\end{align} 
for the \emph{initial force-relaxation rate} (IFRR)
\begin{subequations}
	\begin{align}
	{\bf R}_{\boldsymbol{\lambda}_{\rm i}}[\boldsymbol{\lambda}(t)] &\equiv \int \md x \ {\bf f}(x)\, L[x,\boldsymbol{\lambda}(t)]\, \pi(x|\boldsymbol{\lambda}_{\rm i}) \\
	&= \frac{\md \langle {\bf f} \rangle_{\boldsymbol{\lambda}_{\rm i}}}{\md t}\bigg|_{\boldsymbol{\lambda}(t)} \ ,
	\label{Rate Function}
	\end{align}
\end{subequations}
the rate of change of the conjugate forces at the current control-parameter values (averaged over the initial equilibrium distribution).

Within this approximation, the average excess work is
\begin{align}
\langle W_{\rm ex} \rangle_{\Lambda}
\approx \langle W_{\rm ex} \rangle_{\boldsymbol{\lambda}_{\rm i}} -  \int_{0}^{\Delta t}\md t \, \frac{\md \boldsymbol{\lambda}^{\rm T}}{\md t}\int_{0}^{t}\md t'\, {\bf R}_{\boldsymbol{\lambda}_{\rm i}}[\boldsymbol{\lambda}(t')]\ .
\label{Excess work approx 1}
\end{align}
The first RHS term is the excess work during an instantaneous jump between the initial and final control-parameter values, which equals the \emph{relative entropy} $k_{\rm B} TD(\pi_{\rm i}||\pi_{\rm f}) \equiv k_{\rm B} T\int \md x ~\pi_{\rm i}\ln[\pi_{\rm i}/\pi_{\rm f}]$ between the initial $\pi_{\rm i} \equiv \pi(x|\boldsymbol{\lambda}_{\rm i})$ and final $\pi_{\rm f}\equiv \pi(x|\boldsymbol{\lambda}_{\rm f})$ equilibrium probability distributions~\cite{Large2019}. Integrating \eqref{Excess work approx 1} by parts gives our main theoretical result: for sufficiently short duration, the excess work is
\begin{align}
\langle W_{\rm ex} \rangle_{\Lambda} \approx k_{\rm B} TD(\pi_{\rm i}||\pi_{\rm f})- \int_{0}^{\Delta t}\md t  \, {\bf R}_{\boldsymbol{\lambda}_{\rm i}}^{\rm T}[\boldsymbol{\lambda}(t)] \, 
[\boldsymbol{\lambda}_{{\rm f}} - \boldsymbol{\lambda}(t)] \ .
\label{Excess work approx 2}
\end{align}
The second RHS term is the first-order correction in $\Delta t$, an approximation for the saved work $W_{\rm save}\equiv k_{\rm B}TD(\pi_{\rm i}||\pi_{\rm f}) -W_{\rm ex}$ compared to an instantaneous protocol. {\SB We emphasize that these results stem from the short-time approximation of \eqref{Probability expansion} and do not involve any linear-response approximation. Rather than the small-force and long-duration approximations typical of linear-response and steady-state frameworks~\cite{Ruelle1998,Colangeli2011,Marconi2008}, we make no direct assumptions on the strength of driving and instead assume a short duration so that the probability distribution remains near the initial equilibrium distribution.}

\emph{Initial Force-Relaxation Rate}.---The IFRR can be intuitively understood by considering one-dimensional exponential relaxation. For a discrete jump from initial control-parameter value $\lambda_{\rm i}$ to an intermediate value $\lambda$, an exponentially relaxing mean conjugate force obeys
\begin{align}
\langle f(t) \rangle_{\Lambda} = \langle f \rangle_{\lambda_{\rm i}} + \left(\langle f\rangle_{\lambda}- \langle f \rangle_{\lambda_{\rm i}}\right)e^{-t/\tau(\lambda)} \ ,
\label{exponential relaxing force}
\end{align}
where $\tau(\lambda)$ is the relaxation time of the conjugate force. The IFRR is the initial slope of the mean conjugate force as it relaxes towards equilibrium~\eqref{Rate Function}, which for exponential relaxation is 
\begin{align}
{R}_{{\lambda}_{\rm i}}(\lambda) = \frac{\langle f \rangle_{\lambda_{\rm i}}-\langle f \rangle_{\lambda}}{\tau(\lambda)} \ .
\end{align}
Under simple exponential relaxation, $\tau(\lambda)$ is the same relaxation time defined in Ref.~\onlinecite{OptimalPaths} for slow protocols, thereby connecting short- and long-duration control.

For a small control-parameter jump $\lambda - \lambda_{\rm i}$, static linear-response theory, $\langle f \rangle_{\lambda_{\rm i}}-\langle f \rangle_{\lambda}\approx \beta (\lambda - \lambda_{\rm i})\langle \delta f^2 \rangle_{\lambda_{\rm i}}$, implies that the IFRR further simplifies to
\begin{align}
{R}_{{\lambda}_{\rm i}}(\lambda) \approx \frac{\langle \delta f^2 \rangle_{\lambda_{\rm i}} (\lambda - \lambda_{\rm i})}{\tau(\lambda)} \ .
\label{exponential IFRR} 
\end{align}
The relaxation rate is zero at the initial control-parameter value and increases with larger control-parameter jumps which drive the system further from equilibrium.

\emph{Minimum-dissipation protocols}.---Equation~\eqref{Excess work approx 2} allows for relatively straightforward optimization to determine the \emph{short-time efficient protocol} (STEP), satisfying the Euler-Lagrange equation 
\begin{align}
&\frac{\partial }{\partial \boldsymbol{\lambda}}\left[{\bf R}_{\boldsymbol{\lambda}_{\rm i}}^{\rm T}\left(\boldsymbol{\lambda}\right)\left(\boldsymbol{\lambda}_{{\rm f}}-\boldsymbol{\lambda}^{\rm STEP}\right)\right]\bigg|_{\boldsymbol{\lambda}^{\rm STEP} }= {\bf R}_{\boldsymbol{\lambda}_{\rm i}}\left(\boldsymbol{\lambda}^{\rm STEP}\right) \ .
\label{Euler Lagrange 1}
\end{align}
As an algebraic equation, the solution is a point in control-parameter space, thus the STEP consists of two jumps: a jump at the start from its initial value to the optimal value $\boldsymbol{\lambda}^{\rm STEP}$, and at the end from the optimal value to the final value. The STEP is a jump protocol provided the time-evolution operator $L$ is independent of time derivatives of the control parameters. For Fokker-Planck dynamics this is satisfied if the system is driven by a (generally time-dependent) scalar potential.

To illustrate the two-step minimum-dissipation protocol we have calculated the STEP for diverse model systems (Fig.~\ref{STEP protocol}). In the translating- and breathing-trap systems described by Fokker-Planck dynamics (Supplemental Material~\ref{Harmonic Trap}~\cite{SM}), the STEP jumps halfway between the two endpoints, consistent with the results of Ref.~\onlinecite{Schmiedl2007}. The single-spin-flip and two-state binding/unbinding reaction systems are described by master-equation dynamics (Supplemental Material~\ref{Chemical Reaction} and \ref{Ising Model}~\cite{SM}), with STEPs that jump to points that are respectively larger and smaller than halfway between the initial and final control-parameter values. Specific jump sizes for the STEP depend on the functional form of the IFRR, but the minimum-dissipation protocol always consists of jumps to and from an intermediate control-parameter value. 

The STEP jumps to the point in control-parameter space that maximizes the \emph{short-time power savings} 
\begin{align}
P_{\rm save}^{\rm st}(\boldsymbol{\lambda}) \equiv {\bf R}_{\boldsymbol{\lambda}_{\rm i}}^{\rm T}(\boldsymbol{\lambda})(\boldsymbol{\lambda}_{{\rm f}} - \boldsymbol{\lambda})
\label{eq:power_savings}
\end{align}
due to relaxation at intermediate $\boldsymbol{\lambda}$. The STEP balances fast relaxation rate ${\bf R}_{\boldsymbol{\lambda}_{\rm i}}$ with large final jump $\boldsymbol{\lambda}_{{\rm f}} - \boldsymbol{\lambda}$. The STEP spends the duration $\Delta t$ relaxing at $\boldsymbol{\lambda}^{\rm STEP}$, so for short duration $P_{\rm save}^{\rm st}(\boldsymbol{\lambda}^{\rm{STEP}}) \Delta t$ is the work saved relative to an instantaneous protocol. 

\begin{figure}
	\includegraphics[width=\linewidth]{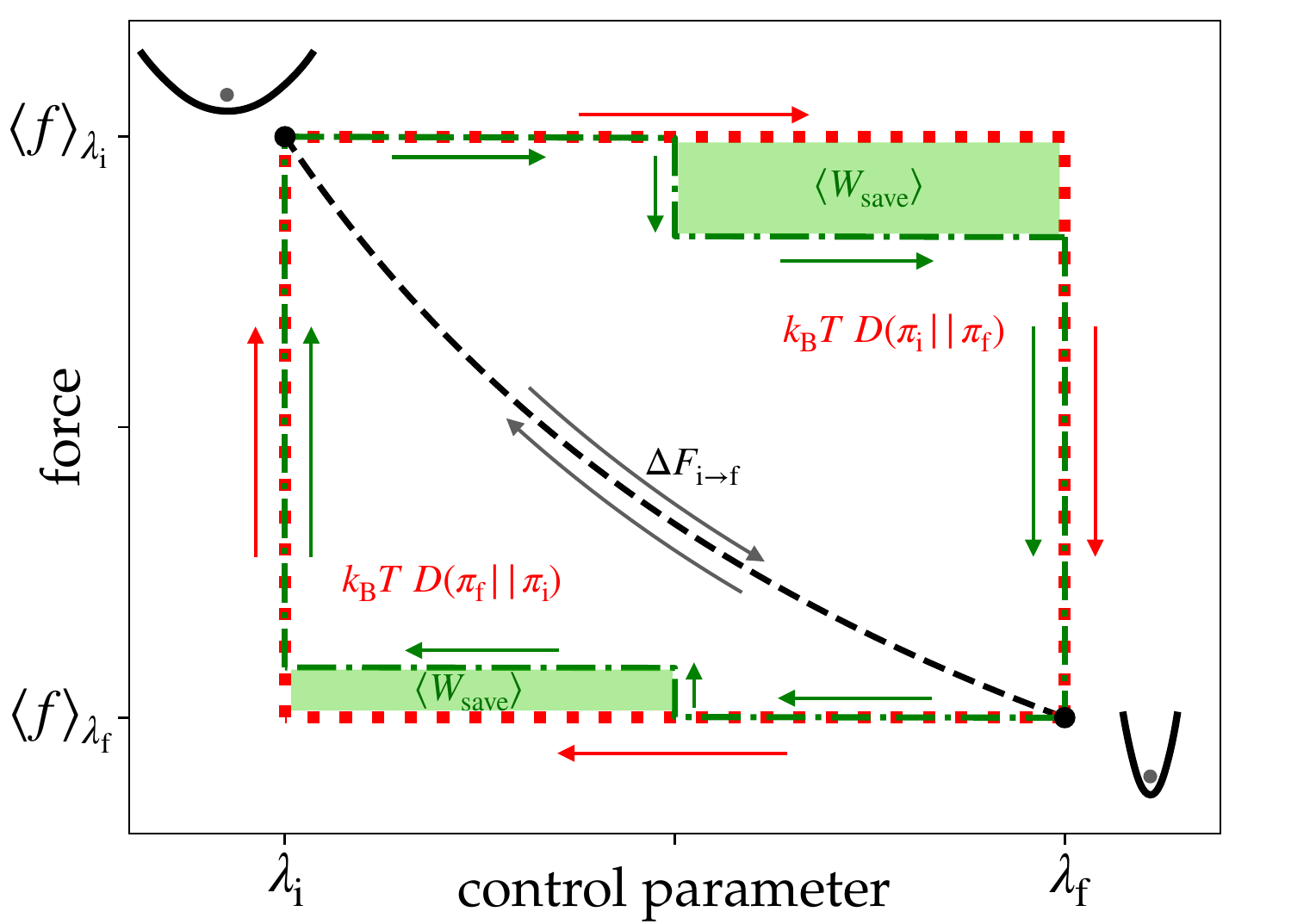}
	\caption{Thermodynamic cycle in the force vs.\ control parameter plane for the breathing harmonic trap driven by instantaneous (red dotted), STEP (green dash-dotted), and quasistatic (black dashed) protocols. Arrows denote protocol direction for transitions between open ($\lambda_{\rm i}$) and closed ($\lambda_{\rm f}$) states, shown schematically. The area under each curve gives the average work done by the respective protocol, the area under the quasistatic curve is the free-energy difference $\Delta F_{{\rm i} \rightarrow {\rm f}} = F_{\rm f} -F_{\rm i}$, the area between the instantaneous (dotted) and quasistatic (dashed) curves is the relative entropy (e.g., $\langle W \rangle_{\lambda_{\rm i}}\ -\Delta F_{{\rm i} \rightarrow {\rm f}}= k_{\rm B} T D(\pi_{\rm i}||\pi_{\rm f})$), and the area between the STEP (dash-dotted) and instantaneous (dotted) curves is the saved work $\langle W_{\rm save}\rangle$ (shaded rectangles). Control-parameter endpoints satisfy $\lambda_{\rm i}/\lambda_{\rm f} = 1/2$, with duration $\Delta t/\tau = 2/5$ for fastest relaxation time $\tau = 1/(2\lambda_{\rm f})$.}
	\label{BHT_Work_Cycle}
\end{figure}

To demonstrate the energetics of the STEP, consider the thermodynamic cycle consisting of tightening and loosening a harmonic trap (Fig.~\ref{BHT_Work_Cycle}). For a quasistatic (infinitely slow) protocol, work equals the free-energy difference, which exactly cancels for a cyclic process. An instantaneous protocol has an additional contribution, which for tightening (loosening) the trap equals the relative entropy between the open (closed) and closed (open) states. The relative entropy is dissipated as heat during equilibration between tightening and loosening the trap (outer vertical arrows). The STEP spends the duration relaxing at an intermediate control-parameter value, resulting in saved work approximated by the area of the rectangle with width given by the final jump size $\lambda_{{\rm f}} - \lambda^{\rm STEP}$ and height by ${R}_{{\lambda}_{\rm i}}(\lambda^{\rm STEP})\Delta t$. To maximize the saved work (rectangle area) the STEP optimally balances the IFRR (determining the height) with final jump size (width). 

For a single control parameter, if the duration is sufficiently short the \emph{gain} $G_{\rm save} \equiv {\langle W_{\rm save} \rangle_{\Lambda}^{\rm des}} /{\langle W_{\rm save} \rangle_{\Lambda}^{\rm naive}}$ in saved work by the STEP is 
\begin{align}
	G_{\rm save}^{\rm STEP} \approx \frac{\max_{\boldsymbol{\lambda}}[P_{\rm save}^{\rm st}(\lambda)]}{\overline{P_{\rm save}^{\rm st}(\lambda)}}\label{Fast_Ratio_Appx} \ ,
\end{align}
where an overbar denotes the spatial average $\overline{P_{\rm save}^{\rm st}(\lambda)} \equiv (\Delta \lambda)^{-1}\int_{\lambda_{\rm i}}^{\lambda_{\rm f}}\md \lambda~P_{\rm save}^{\rm st}(\lambda)$, ``naive'' the constant-velocity protocol, and ``des'' a designed protocol. The gain from a STEP is greatest when the power savings $P_{\rm save}^{\rm st}(\lambda)$ is sharply peaked.

\emph{Interpolated protocols}.---In order to design a protocol that performs well for any duration, we combine the STEP (optimal for fast protocols) with the minimum-dissipation protocol from linear-response theory (optimal for slow protocols). The linear-response protocol is continuous and when driven by a single control parameter proceeds at velocity $\md \lambda/\md t \propto [\zeta(\lambda)]^{-1/2}$, where $\zeta(\lambda)$ is the generalized friction coefficient~\cite{OptimalPaths}. We assume the shape of the protocol from linear-response theory remains unchanged (i.e., $\md \lambda/\md t \propto [\zeta(\lambda)]^{-1/2}$) but with initial jump $(\lambda^{\rm STEP} -\lambda_{\rm i})/(1+\Delta t/\tau)^{\alpha}$ and final jump $(\lambda_{\rm f}-\lambda^{\rm STEP})/(1+\Delta t/\tau)^{\alpha}$, where the constant $\alpha$ controls the crossover from slow to fast approximations. For our simple systems we empirically find $\alpha =1$ performs relatively well.
 
Figure~\ref{BHT_Work_Diff} shows the improvement from designed protocols relative to naive (constant-velocity) for the model system of a breathing harmonic trap. The difference between naive and designed work (Fig.~\ref{BHT_Work_Diff}a) shows the expected asymptotic behavior in the short- and long-duration limits: scaling as $\Delta t$ (slope of one) for small $\Delta t/\tau$ and $(\Delta t)^{-1}$ (slope of negative one) for large $\Delta t/\tau$. Both the fast and slow designed protocols perform worse than naive (the difference is negative) for large and small $\Delta t/\tau$, respectively. The fast-protocol approximation~\eqref{Excess work approx 2} breaks down for long duration because the conjugate-force relaxation rate decreases as the system approaches equilibrium at $\boldsymbol{\lambda}$, whereas \eqref{Excess work approx 2} assumes a constant relaxation time. However, the interpolated protocol performs well for any duration, and the largest work saved compared to naive is for intermediate duration. The gain $G_{\rm save}$ quantifies the percent increase in saved work from designed protocols relative to naive, where a gain greater than one indicates the designed does less work than the naive. For this system, the largest gain in saved work occurs in the fast limit (small $\Delta t/\tau$) for the STEP, interpolated, and exact optimal protocols. 

\emph{Multi-dimensional control}.---Optimization of multi-dimensional control protocols has seen a recent surge in interest, primarily driven by possible improvements to nonequilibrium free-energy estimates in fast-switching simulations~\cite{Chipot2011,Dellago2014}. Previous calculations of minimum-dissipation protocols which observed jumps were limited to one-dimensional optimization. A significant advantage of the present approximation is that it permits simple multidimensional control-protocol optimization. Equation~\eqref{Euler Lagrange 1} implies that for multidimensional control the STEP consists of jumps to and from the control-parameter point $\boldsymbol{\lambda}^{\rm STEP}$.

To illustrate, we consider a nine-spin Ising model with frustrated boundary conditions (Fig.~\ref{MultiD_Ising})~\cite{Rotskoff2017,Venturoli2009}. We use a 2D control parameter $\boldsymbol{h} = (h_{\rm b},h_{\rm g})$ of magnetic fields applied to the mid-edge spins (Fig.~\ref{MultiD_Ising}a) which initially hold the system in the spin-down state and reverse during the protocol, driving the system to invert its magnetization. Supplemental Material~\ref{multiD_Ising_Model}~\cite{SM} gives model details.

The power saving~\eqref{eq:power_savings} vanishes at initial and final control-parameter values, respectively corresponding to zero relaxation rate and zero final jump size (Fig.~\ref{MultiD_Ising}b). By jumping past control-parameter regions with small power saving, the STEP outperforms the naive protocol for short duration, as quantified by the difference between naive and designed work (Fig.~\ref{MultiD_Ising}c) and the gain in saved work (Fig.~\ref{MultiD_Ising}d). Indeed, for short duration the STEP more than doubles the work saved by the naive protocol (i.e., has gain greater than two).

\emph{Discussion}.---We have developed an approximation for work in the fast-protocol limit~\eqref{Excess work approx 2} that permits straightforward optimization~\eqref{Euler Lagrange 1} simply from the initial force-relaxation rate (IFRR), Eq.~\eqref{Rate Function}. We find that jumps are a universal feature of minimum-dissipation protocols in this fast limit, which we illustrate with several model systems under single- (Fig.~\ref{STEP protocol}) or multi-dimensional control (Fig.~\ref{MultiD_Ising}). Jumps minimize dissipation for fast protocols because the relaxation rate is approximately constant, with no diminishing returns from spending the entire duration at a single control-parameter value. Therefore, the STEP jumps between the fixed control-parameter endpoints to spend the entire duration at the control-parameter value that maximizes the product of the IFRR and the subsequent jump size~\eqref{eq:power_savings}. This breaks down for slow protocols since with sufficient time at a single control-parameter value, the relaxation rate decreases over time; indeed, in the slow limit the minimum-dissipation protocol is continuous~\cite{OptimalPaths}. We combine these two seemingly disparate limits with a simple interpolation scheme, producing protocols that perform well for any duration (Fig.~\ref{BHT_Work_Diff}).

One important application of minimum-dissipation protocols is to free-energy estimation, which aids the design of novel ligands for targeted protein binding~\cite{Schindler2020,Aldeghi2018,Kuhn2017,Ciordia2016,Wang2015,Gapsys2015,Chodera2011}. Quite generally, the accuracy of free-energy estimates decreases with increasing dissipation~\cite{Ritort2002,Gore2003,Shenfeld2009,Geiger2010,Kim2012,Blaber2020Skewed}. Based on the results of Ref.~\onlinecite{Schmiedl2007}, jump protocols have been used to reduce dissipation and improve free-energy estimates~\cite{Geiger2010}, but previously it was unknown whether jumps would always reduce dissipation in these more complex systems, and there was no simple procedure to find the optimal jump size. The present formalism demonstrates that jumps are a general feature and gives a simple optimization procedure applicable to multidimensional control. This makes protocol optimization tractable for a considerably expanded range of systems.

Although we focused on systems with known equations of motion, the IFRR~\eqref{Rate Function} and short-time power savings~\eqref{eq:power_savings} are easily estimated without detailed dynamical information: the system only needs to be equilibrated at a single control-parameter value; the protocol can be very short; the average converges with few samples; and the STEP is found using standard optimization techniques applied to \eqref{eq:power_savings}. The STEP can thus be computed relatively inexpensively, easing determination of minimum-dissipation protocols in rapidly driven complex chemical and biological systems. This opens the door to improve the energetic efficiency in thermodynamic computing~\cite{Conte2019,Proesmans2020} and the accuracy of nonequilibrium free-energy estimates in simulations and single-molecule experiments~\cite{Tafoya2019,Blaber2020Skewed,Shenfeld2009,Ritort2002}.

\vspace{2ex}
This work is supported by an SFU Graduate Deans Entrance Scholarship (SB), an NSERC CGS Doctoral fellowship (MDL), an NSERC Discovery Grant and Discovery Accelerator Supplement (DAS), and a Tier-II Canada Research Chair (DAS), and was enabled in part by support provided by WestGrid (www.westgrid.ca) and Compute Canada Calcul Canada (www.computecanada.ca). The authors thank John Bechhoefer, Jannik Ehrich, and Joseph Lucero (SFU Physics) for enlightening feedback on the manuscript.

\onecolumngrid
\clearpage
\begin{center}
	\textbf{\large Supplemental material for ``Steps minimize dissipation in rapidly driven stochastic systems''}
\end{center}
\setcounter{equation}{0}
\setcounter{figure}{0}
\setcounter{table}{0}
\setcounter{page}{1}
\makeatletter
\renewcommand{\theequation}{S\arabic{equation}}
\renewcommand{\thefigure}{S\arabic{figure}}

\section{Harmonic trap}
\label{Harmonic Trap}
We consider a colloidal particle in a harmonic trap with energy $U = \frac{1}{2}k(t)[x-u(t)]^2$, for variable trap center $u(t)$ and trap stiffness $k(t)$. The particle position $x$ obeys the overdamped Langevin equation,
\begin{align}
\gamma\frac{\md x}{\md t} = -k(t)[x-u(t)] +\sqrt{2\gamma k_{\rm B}T} \, \eta \ ,
\end{align}
for damping coefficient $\gamma$, temperature $T$, and Gaussian white noise $\eta$. For simplicity we choose units where $\gamma = k_{\rm B}T = 1$, and scale the trap stiffness by its initial value $k_{\rm i}$. The corresponding Fokker-Planck equation is
\begin{align}
\frac{\partial p_{\Lambda}(x,t)}{\partial t} = L[x,\{k(t),u(t)\}]\, p_{\Lambda}(x,t) \ ,
\label{Harmonic FP}
\end{align}
with
\begin{align}
L[x,\{k(t),u(t)\}] \equiv \frac{\partial }{\partial x}\left[k(t)(x-u(t))+\frac{\partial }{\partial x}\right] \ .
\end{align}
For this system, the two control parameters $u(t)$ and $k(t)$ can be treated independently as ``translating-trap'' and ``breathing-trap'' models respectively.

\subsection{Translating trap}
\label{Translating trap}

In a translating trap we hold the trap stiffness $k$ fixed and vary the trap center $u(t)$ as the control parameter. The conjugate force is therefore $f \equiv -\partial U/\partial u = k[x-u(t)]$, and the IFRR~\eqref{Rate Function} is
\begin{align}
{R}_{{u}_{\rm i}}[u(t)] = k^2[u_{\rm i}-u(t)] \ ,
\end{align}
for initial trap center $u_{\rm i}$. Solving \eqref{Euler Lagrange 1} gives the STEP value
\begin{align}
u^{\rm STEP} = \frac{u_{\rm i}+u_{\rm f}}{2} \ ,
\end{align}
for final trap center $u_{\rm f}$. The STEP jumps the trap center halfway between its initial and final positions, independent of any other system parameters. This is consistent with the exact minimum-dissipation protocol in the fast (short-duration) limit~\cite{Schmiedl2007}.

\subsection{Breathing trap}
\label{Breathing trap}

The breathing trap has fixed trap center ($u = 0$) and time-dependent trap stiffness $k(t)$ as the control parameter. Here the conjugate force is $f \equiv -\partial U/\partial k = -\frac{1}{2}x^2$, so the IFRR~\eqref{Rate Function} is
\begin{align}
{R}_{{k}_{\rm i}}[k(t)] = 2\left[1-\frac{k(t)}{k_{\rm i}}\right] \ ,
\end{align}
for initial stiffness $k_{\rm i}$. Solving \eqref{Euler Lagrange 1} gives the STEP value
\begin{align}
k^{\rm STEP}= \frac{k_{\rm i}+k_{\rm f}}{2} \ ,
\end{align}
where $k_{\rm f}$ is the final trap stiffness. Identical to the translating trap, the STEP jumps to the control-parameter value halfway between the endpoints, independent of other system parameters, consistent with the exact result in the fast (short-duration) limit~\cite{Schmiedl2007}. The approximate gain~\eqref{Fast_Ratio_Appx} from the STEP is $3/2$, independent of system parameters.

For a slow (long-duration) protocol, the minimum-dissipation protocol is continuous, and can be calculated from the friction coefficient~\cite{OptimalPaths}
\begin{align}
\zeta(k) = \frac{1}{4k^3} \ ,
\end{align}
as $\md k/\md t \propto [\zeta(k)]^{-1/2}$, where the proportionality is set by the control-parameter endpoints. For the interpolated protocol, we add jumps at the beginning and end of size $(k^{\rm STEP} - k_{\rm i})/(1+\Delta t/\tau)^{\alpha}$ and $(k_{\rm f} - k^{\rm STEP})/(1+\Delta t/\tau)^{\alpha}$ respectively.

\section{Binding and unbinding reaction}
\label{Chemical Reaction}
We examine a two-state binding/unbinding reaction with binding rate $k_{\rm UB \rightarrow B}$ and unbinding rate $k_{\rm B \rightarrow UB}$~\cite{Blaber2020}. We assume the binding rate $k_{\rm UB \rightarrow B}=k_{0}$ depends only on the dynamic encounter rate and not on the strength of the chemical potential, and the unbinding rate $k_{\rm B \rightarrow UB}$ depends on how tightly the molecule is bound, and hence on the chemical-potential difference $\mu$ between unbound and bound states, as (with $\beta = 1$)
\begin{equation}
\label{rate chemical dependence}
k_{\rm B \rightarrow UB} = k_{0}e^{-\mu} \ .
\end{equation}
$\mu = 0$ gives equal binding and unbinding rates, $k_{\rm UB \rightarrow B} = k_{\rm B \rightarrow UB}$.

We additionally assume a fixed total number $\Ntot = \Nub+\Nb$ of molecules, with variable numbers of unbound ($\Nub$) and bound ($\Nb$) molecules. The transition-rate matrix is
\begin{align}
K(\mu) =	\left[ {\begin{array}{cc}
	k_{0}e^{- \mu}&-k_{0} \\
	-k_{0}e^{- \mu}&k_{0} \\
	\end{array} } \right] \ .
\end{align}
The excess work for this two-state system driven by a chemical-potential protocol can be solved by numerically integrating
\begin{align}
\frac{\md \langle \Nb \rangle_{\Lambda}}{\md t} = k_{0}\left[\Ntot -\langle \Nb \rangle_{\Lambda} (e^{-\mu(t)}+1)\right] \ ,
\label{changeinN}
\end{align}
subject to an equilibrium initial condition
\begin{align}
\langle \Nb \rangle_{\mu} = \frac{\Ntot}{1+e^{-\mu}} \ .
\end{align}

Using \eqref{changeinN} gives the IFRR~\eqref{Rate Function} 
\begin{align}
{R}_{{\mu}_{\rm i}}[\mu(t)] = -k_{0}\left[\Ntot -\langle \Nb \rangle_{\muI} (e^{-\mu(t)}+1)\right] \ .
\end{align}
Solving \eqref{Euler Lagrange 1} gives the STEP value
\begin{align}
\label{CHR fast protocol}
\mu^{\rm STEP} = \mathcal{W}\left[\left(1-\frac{\Ntot}{\langle \Nb \rangle_{\muI}}\right)e^{(\muF-1)}\right] +\muF-1 \ ,
\end{align}
where $\mathcal{W}$ is the product log function (Lambert W function), defined as the solution to $\mathcal{W}(z)\exp[\mathcal{W}(z)]=z$. For the special case of $\langle \Nb \rangle_{\muI} \approx \Ntot$ (satisfied as $\muI\rightarrow\infty$), the STEP value simplifies to $\mu^{\rm STEP} \approx \muF-1$. Figure~\ref{STEP protocol} shows the STEP for $\muI = -3+\ln2$ and $\muF = 3+\ln2$.

In \cite{Blaber2020}, we derived simple expressions for the friction
\begin{align}
\label{CHR friction}
\zeta(\mu) &= \frac{\Ntot}{k_{0}} \frac{e^{-\mu}}{(1+e^{-\mu})^3} \ ,
\end{align}
and minimum-dissipation protocol in the slow limit,
\begin{align}
\label{CHR slow protocol}
\frac{\md \mu(t) }{\md t} = \frac{e^{-\mu(t)}\left(e^{\muF}-e^{\muI} \right) }{\Delta t} \ .
\end{align}
From \eqref{CHR fast protocol} we create an interpolated protocol which satisfies~\eqref{CHR slow protocol} but with initial jump $(\mu^{\rm STEP} - \mu_{\rm i})/(1+\Delta t/\tau)^{\alpha}$ and final jump $(\mu_{\rm f} - \mu^{\rm STEP})/(1+\Delta t/\tau)^{\alpha}$. 

\begin{figure}
	\includegraphics[width=0.5\textwidth]{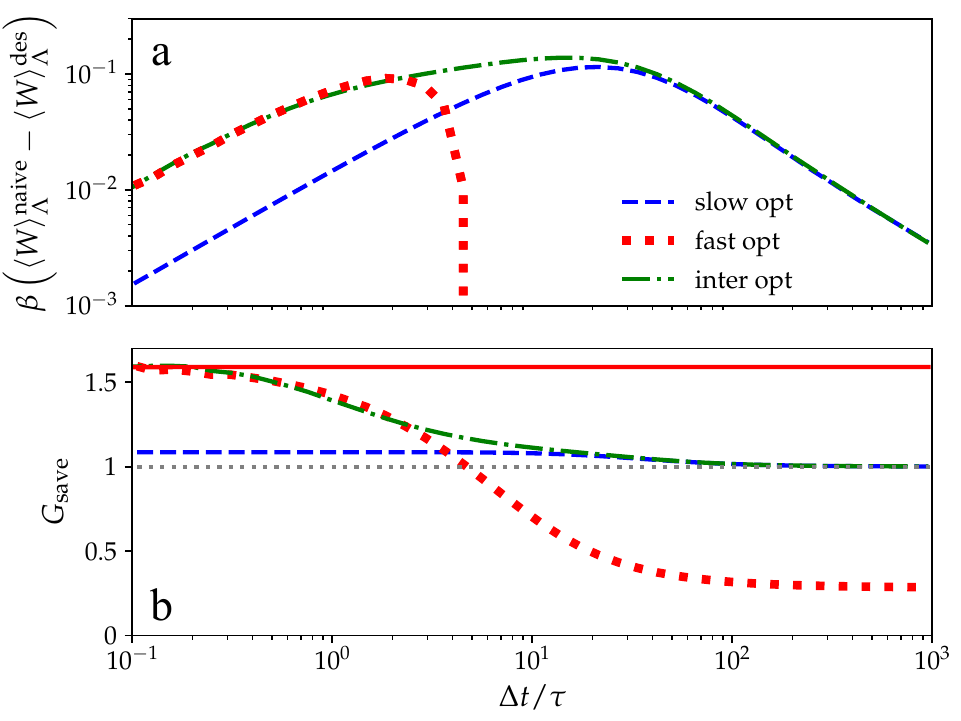}
	\caption{Benefit in the two-state binding/unbinding reaction system from designed protocols relative to naive (constant-velocity), as a function of total duration $\Delta t$ scaled by the fastest integral relaxation time $\tau$. The different designed (``des'') protocols include the linear-response optimized (``slow opt'', dashed blue), STEP (``fast opt'', red dots), and interpolated optimal protocol with $\alpha=1$ (``inter opt'', dash-dotted green). (a) Difference between the work in naive (constant-velocity) and designed protocols. (b) Gain $G_{\rm save}\equiv {\langle W_{\rm save} \rangle_{\Lambda}^{\rm des}} /{\langle W_{\rm save} \rangle_{\Lambda}^{\rm naive}}$ in saved work, with a solid red line denoting the short-duration limit~\eqref{Fast_Ratio_Appx}. Control-parameter endpoints are $ \muI = -3+\ln 2$ and $\muF = 3+\ln 2$, and the fastest integral relaxation time is $\tau = 1/[k_{0}(1+e^{-\mu_{\rm i}})]$~\cite{Blaber2020}.}
	\label{CHR_Work_Diff}
\end{figure}

Figure~\ref{CHR_Work_Diff} shows the benefit from designed protocols compared to naive (constant-velocity) protocols. Consistent with the breathing trap (Fig.~\ref{BHT_Work_Diff}), the difference between the naive and designed work in Fig.~\ref{CHR_Work_Diff} demonstrates the expected $(\Delta t)^{-1}$ scaling of work in the slow (long-duration) limit, $(\Delta t)$ scaling in the fast (short-duration) limit, and the largest work reduction from designed protocols in the intermediate ($\Delta t/\tau \sim 10$) regime, achieved by the interpolated protocol. Unlike the breathing trap, for any duration the protocol designed from slow approximations performs better than the naive (positive difference in Fig.~\ref{CHR_Work_Diff}a), but can still be significantly outperformed by a protocol incorporating the fast approximation, as shown by the larger gain from ``fast opt'' compared to ``slow opt'' for short protocol duration in Fig.~\ref{CHR_Work_Diff}b. 

\section{Single-spin Ising model}
\label{Ising Model}

Consider a single-spin Ising model under the control of an external magnetic field $h$ with Hamiltonian $H(\sigma|h) = -h \sigma$, where $\sigma \in \{-1,1\}$. The system dynamically evolves according to a master equation with transition-rate matrix
\begin{equation}
K(h) = 
\begin{bmatrix}
-k_{1}(h)  & k_{-1}(h) \\
k_{1}(h)  & -k_{-1}(h) \ ,
\end{bmatrix}
\label{Ising Master Equation}
\end{equation}
for rates
\begin{subequations}
\begin{align}
k_{1}(h) &= k_0 \frac{1}{1+e^{-2\beta h}} \\
k_{-1}(h) &= k_0 \frac{1}{1+e^{2\beta h}} \ ,
\label{Ising Reaction Rates} 
\end{align}
\end{subequations}
where $k_0$ is the rate of spin-flip attempts and the second factor is the Glauber acceptance probability~\cite{Glauber1963}. The IFRR~\eqref{Rate Function} is
\begin{subequations}
\label{Ising Rate Function}
\begin{align}
{R}_{{h}_{\rm i}}[h(t)] &= \sum_{\sigma \in \{-1,1\}} \sigma K(h(t)) \pi_{\rm i} \\
&= k_0 \sech \beta h \ \sech \beta h_{\rm i}  \ \sinh{\beta(h-h_{\rm i})} \ ,
\end{align}
\end{subequations}
where $h_{\rm i}$ is the initial magnetic field. The STEP value~\eqref{Euler Lagrange 1} is found by solving the transcendental equation
\begin{equation}
    \left[\coth \beta (h^{\rm STEP}-h_{\rm i})
    -\tanh \beta h^{\rm STEP} \right]\left(h_{\rm f}-h^{\rm STEP}\right) = 1 \ .
\end{equation}

From Ref.~\onlinecite{OptimalPaths}, the generalized friction coefficient $\zeta(h)$ is the product of the equilibrium conjugate-force variance $\langle \delta \sigma^2 \rangle_{h}$ and integral relaxation time $\tau(h)=[k_{1}(h)+k_{-1}(h)]^{-1}=k_0^{-1}$, giving
\begin{equation}
    \zeta(h) = \beta k_0^{-1} \sech^2 \beta h \ .
\end{equation}
(The relaxation time is derived from the second eigenvalue of the transition-rate matrix~\eqref{Ising Master Equation}.) In the long-duration limit, this yields the minimum-dissipation protocol 
\begin{equation}
    \frac{\md h}{\md t} = \frac{2 \kT}{\Delta t} \left[ g \left( \tfrac{1}{2}\beta h_{\rm f} \right) - g \left( \tfrac{1}{2}\beta h_{\rm i} \right) \right] \cosh \beta h  \ ,
\end{equation}
where $g(x)\equiv\tan^{-1} [\tanh x]$.

\begin{figure}
    \centering
    \includegraphics[width=0.5\textwidth]{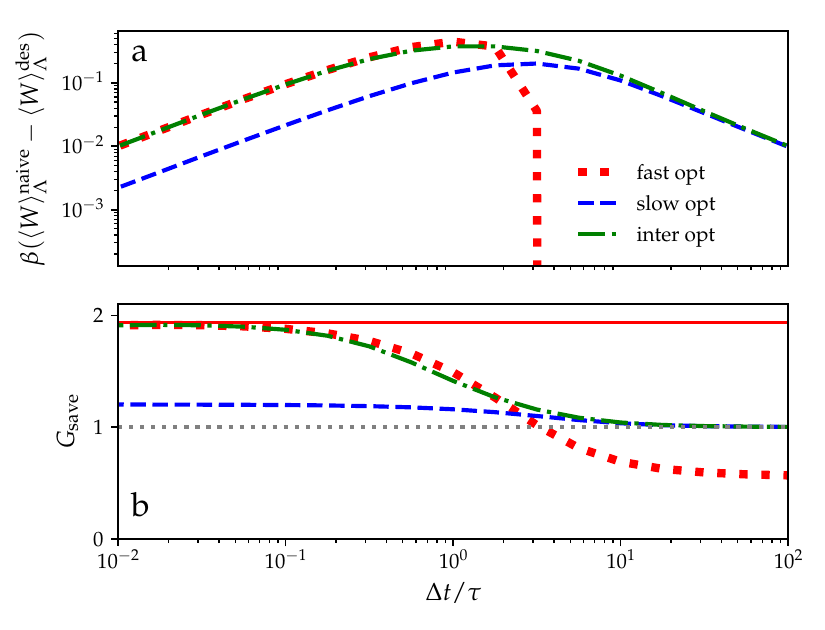}
    \caption{Benefit in the single-spin Ising model from designed protocols relative to the naive (constant-velocity) protocol, as a function of protocol duration $\Delta t$ scaled by the integral relaxation time $\tau = k_0^{-1}$. The different designed (``des'') protocols include the linear-response optimized (``slow opt'', dashed blue), STEP (``fast opt'', red dots), and the interpolated optimal protocol with $\alpha=1$ (``inter opt'', dash-dotted green). (a) Difference between the work by naive (constant-velocity) and by designed (``des'') protocols. (b) Gain $G_{\rm save}\equiv {\langle W_{\rm save} \rangle_{\Lambda}^{\rm des}} /{\langle W_{\rm save} \rangle_{\Lambda}^{\rm naive}}$ in saved work, with a solid red line denoting the short-duration limit~\eqref{Fast_Ratio_Appx}. Control-parameter endpoints are $\beta h_{\rm i} = -2$ and $\beta h_{\rm f} = 2$.}
    \label{fig:Ising1D}
\end{figure}

Figure~\ref{fig:Ising1D} shows the benefit from designed protocols compared to naive protocols. The results are qualitatively similar to the binding/unbinding reaction system; both are two-state systems with a control parameter that biases the transition rate between the two states, but the rate matrices have different analytical forms.

\section{Multi-dimensional control of nine-spin Ising model}
\label{multiD_Ising_Model}

We consider the nine-spin ferromagnetic Ising model depicted in Fig.~\ref{MultiD_Ising}. Adjacent spins interact with coupling strength $\beta J=0.5$, both to the fluctuating spins and fixed-spin boundary conditions. The forces $\bf{f}$ $=(m_{\rm{b}},m_{\rm{g}})$ conjugate to the external magnetic fields $\boldsymbol{h}=(h_{\rm b},h_{\rm g})$ are the mean magnetizations of the spins controlled by each field. The spins dynamically evolve according to Glauber dynamics~\cite{Glauber1963}.

The IFRR~\eqref{Rate Function} is calculated over the control-parameter space, and the short-time power savings in Fig.~\ref{MultiD_Ising}b is obtained from \eqref{eq:power_savings}. Mean works for the naive and designed protocols (Fig.~\ref{MultiD_Ising}c/d) were calculated by propagating the master equation for protocol durations $\Delta t/\tau$ ranging between $10^{-3}$ and $10^3$, with simulation time-step $\md t \in (2 \times 10^{-4},2 \times 10^{-1})$, using a fine temporal discretization for short protocol durations and a coarse discretization for long protocol durations.

\end{document}